\documentclass[%
pre, reprint, onecolumn, superscriptaddress, longbibliography,
amsmath, amssymb, aps, floatfix,
]{revtex4-2}
\usepackage{bm}
\usepackage{microtype}
\usepackage[colorlinks=true,allcolors=blue]{hyperref}
\usepackage{graphicx}

\begin{document}

\title{Migration Flows: Population Scaling and Heavy Tails}

\author{Cook Hyun Kim}
\thanks{Corresponding author: protozerta@kentech.ac.kr}
\affiliation{Department of Energy Engineering, Korea Institute of Energy Technology, Naju 58330, Korea}
\author{Sangjoon Park}
\affiliation{Department of Energy Engineering, Korea Institute of Energy Technology, Naju 58330, Korea}
\author{Seoyoung Jin}
\affiliation{Department of Transportation and Logistics Engineering, Hanyang University, Ansan 15588, Korea}
\author{Jaiyong Lee}
\affiliation{Department of Energy Engineering, Korea Institute of Energy Technology, Naju 58330, Korea}

\date{\today}

\begin{abstract}
We show how migration between counties in different cities can explain both the population scaling of mean intercity flows and heavy-tailed intercity net-flow distributions. Density-dependent departure and arrival activities determine how mean flows scale with the populations of the origin and destination cities. We assume Gaussian county-pair net flows with standard deviations proportional to their mean directional flows. At a fixed correlation between county-pair net flows, their sum is Gaussian. Averaging this conditional Gaussian density over the correlation distribution can produce an intermediate power law with a Gaussian cutoff. We also present a stochastic model that generates the variance distribution used in this mixture.
\end{abstract}

\maketitle

\section{Introduction}

Migration connects cities and redistributes populations. Gravity models relate migration flows to populations and travel costs~\cite{zipf1946,wilson1967}, while the radiation model relates migration flows to the spatial distribution of opportunities~\cite{simini2012}. Stochastic growth models describe population dynamics through demographic fluctuations~\cite{bettencourt2020}. Comparative studies evaluate migration models~\cite{lenormand2016}, and reviews summarize mobility models and applications~\cite{barbosa2018}. Understanding migration requires examining both mean intercity flows and intercity net-flow distributions.

Verbavatz and Barthelemy~\cite{verbavatz2020} report mean intercity flows scaling as $F_{ij}\propto S_i^\alpha S_j^\alpha$ and heavy-tailed distributions of population-adjusted intercity net flows. Here $S_i$ is the population of city $i$, $J_{i\to j}$ is the directional flow from city $i$ to city $j$, and $F_{ij}=\mathbb E[J_{i\to j}]$ is the mean directional flow. The net flow into city $i$ from city $j$ is $Z_{ij}=J_{j\to i}-J_{i\to j}$. The reported heavy tails motivate a L\'evy-noise description of city growth. Two questions follow: what determines the population scaling of mean intercity flows, and what produces heavy-tailed intercity net-flow distributions?

A simple model assigns each person in city $i$ a migration rate $cS_j$ toward city $j$, giving $F_{ij}=cS_iS_j$. The mean intercity flow is then linear in each city's population. If the intercity net flow sums many independent contributions with finite variances and no dominant contribution, the intercity net-flow distribution is approximately Gaussian. Explaining nonlinear population scaling of mean intercity flows and heavy-tailed intercity net-flow distributions therefore requires additional mechanisms.

Internal population structure offers one possible route. Reia et al.~\cite{reia2022spatial,reia2022modeling} find that intercity migration is concentrated in central counties, while migration within cities tends to redistribute population toward outer counties. Reia et al.\ also report approximately Gaussian distributions of population-adjusted total net flows at the county level. The county-level results motivate asking whether Gaussian county-pair net flows can combine to produce heavy-tailed intercity net-flow distributions.

\begin{figure}[tb]
\centering
\includegraphics[width=0.75\linewidth]{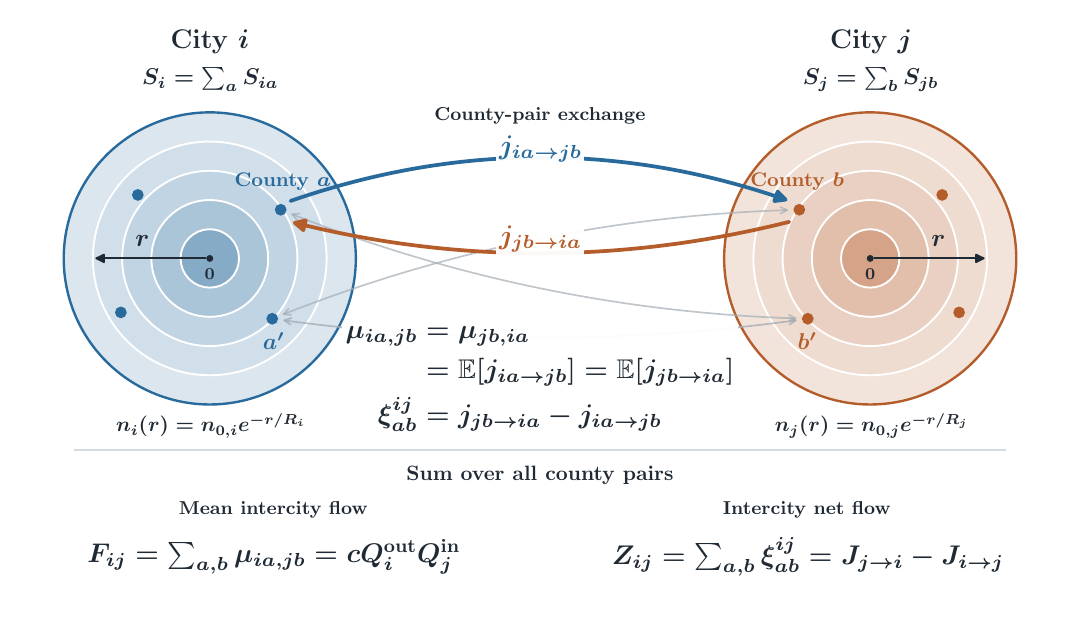}
\caption{Migration between counties in two cities. Concentric rings illustrate radial population-density profiles, with darker shading indicating higher density. Dots represent counties. Colored arrows show opposing flows between counties $a$ and $b$, while gray arrows show other county-pair exchanges. Opposing flows have equal means but can differ in each realization, producing net-flow fluctuations. Summing county-pair mean directional flows gives the mean intercity flow $F_{ij}$, while summing county-pair net flows gives the intercity net flow $Z_{ij}$.}
\label{fig:schematic}
\end{figure}

We represent migration as effective flows between departure and final arrival counties in different cities, without modeling intermediate moves (Fig.~\ref{fig:schematic}). Summing county-pair directional flows gives the intercity directional flow, while summing county-pair net flows gives the intercity net flow. We assume Gaussian county-pair net flows with standard deviations proportional to the corresponding mean directional flows. The Gaussian assumption applies to individual county pairs; the empirical studies examine total net flows at the county level~\cite{reia2022spatial,reia2022modeling}.

The model explains the population scaling of mean intercity flows through density-dependent departure and arrival activities. Integrating local departure activity over the origin city gives the total departure activity, while integrating local arrival activity over the destination city gives the total arrival activity. The mean intercity flow is proportional to the product of the total departure and arrival activities. The scaling of mean intercity flows with city populations therefore depends on how population density changes with city population and how departure and arrival activities change with population density. The resulting population scaling connects the model to nonlinear urban scaling~\cite{bettencourt2007,bettencourt2013}.

The model explains heavy-tailed intercity net-flow distributions through correlations between county-pair net flows. At a fixed correlation, the intercity net flow is Gaussian, with a variance determined by the correlation. Averaging the conditional Gaussian density over the correlation distribution gives a mixture of Gaussian densities with different variances. For a suitable variance distribution and a sufficiently broad variance range, the mixture follows an intermediate power law with a Gaussian cutoff. The averaging over variances connects the model to Gaussian mixtures and superstatistics~\cite{andrews1974,west1987,beck2003}.

County-pair migration thus provides a common framework for explaining both the population scaling of mean intercity flows and heavy-tailed intercity net-flow distributions.

\section{Mean flows from population density}

\subsection{Local density and activity}

We use an exponential population-density profile~\cite{clark1951}:
\begin{equation}
n_i(r)=n_{0,i}e^{-r/R_i},\qquad
S_i=\int_0^\infty n_i(r)\,2\pi r\,dr
=2\pi n_{0,i}R_i^2,
\label{eq:density}
\end{equation}
where $r$ is the distance from the center of city $i$, $n_{0,i}$ is the central population density, and $R_i$ is the characteristic radius. The factor $2\pi r\,dr$ is the area of a thin ring.

We assume that central population density scales with city population as
\begin{equation}
n_{0,i}=n_\star(S_i/S_\star)^\delta,
\label{eq:amplitude}
\end{equation}
where $n_\star$ is a fixed reference density and $S_\star$ is a fixed reference population. Combining Eqs.~\eqref{eq:density} and \eqref{eq:amplitude} gives $R_i\propto S_i^{(1-\delta)/2}$.

We describe local departure and arrival activities by
\begin{equation}
a_i^\ell(r)=\eta_\ell n_{\rm ref}
\left[\frac{n_i(r)}{n_{\rm ref}}\right]^{q_\ell},
\qquad \ell\in\{\mathrm{out},\mathrm{in}\},
\label{eq:local_activity}
\end{equation}
where $n_{\rm ref}$ is a fixed reference density. The activity exponent $q_\ell>0$ and coefficient $\eta_\ell>0$ are common across cities. The departure activity $a_i^{\rm out}$ measures departure opportunities per unit area, while the arrival activity $a_i^{\rm in}$ measures arrival opportunities per unit area. Activity per person scales as $a_i^\ell/n_i\propto n_i^{q_\ell-1}$. Activity per person is independent of density for $q_\ell=1$, increases with density for $q_\ell>1$, and decreases with density for $0<q_\ell<1$.

Integrating local activity over city $i$ gives the total departure or arrival activity:
\begin{equation}
Q_i^\ell=\int_0^\infty a_i^\ell(r)\,2\pi r\,dr
=\frac{\eta_\ell}{q_\ell^2}S_i
\left(\frac{n_{0,i}}{n_{\rm ref}}\right)^{q_\ell-1}.
\label{eq:integrated_activity}
\end{equation}
Substituting Eq.~\eqref{eq:amplitude} into Eq.~\eqref{eq:integrated_activity} gives
\begin{equation}
Q_i^{\rm out}\propto S_i^\alpha,\quad
\alpha=1+\delta(q_{\rm out}-1),\qquad
Q_i^{\rm in}\propto S_i^\beta,\quad
\beta=1+\delta(q_{\rm in}-1).
\label{eq:exponents}
\end{equation}
The exponent $\alpha$ describes how total departure activity scales with city population, while $\beta$ describes how total arrival activity scales with city population. Both exponents depend on how central density scales with city population and how local activity responds to density.

The activity scaling in Eq.~\eqref{eq:exponents} also holds for other population-density profiles with a common shape across cities. Retaining Eq.~\eqref{eq:amplitude}, consider
\begin{equation}
n_i(r)=n_{0,i}f(r/R_i),
\end{equation}
where $f$ is a nonnegative function common to all cities and $f(0)=1$. Central density and characteristic radius may vary across cities, but the rescaled profile remains the same.

Define $I_p=\int_0^\infty u f(u)^p\,du$, where $u=r/R_i$. Integrating population density and local activity gives
\begin{equation}
S_i=2\pi n_{0,i}R_i^2 I_1,\qquad
Q_i^\ell=\eta_\ell\frac{I_{q_\ell}}{I_1}S_i
\left(\frac{n_{0,i}}{n_{\rm ref}}\right)^{q_\ell-1},
\end{equation}
provided $I_1$ and $I_{q_\ell}$ are finite and positive. For the exponential profile, $I_1=1$ and $I_{q_\ell}=1/q_\ell^2$, recovering Eq.~\eqref{eq:integrated_activity}. Changing the common profile changes the prefactor $I_{q_\ell}/I_1$ but leaves the activity scaling exponents $\alpha$ and $\beta$ unchanged.

\subsection{Mean intercity flows and symmetry}

We assume that the mean flow from a small area $dA_{\bm x}$ in city $i$ to a small area $dA_{\bm y}$ in city $j$ is proportional to the departure activity in $dA_{\bm x}$ and the arrival activity in $dA_{\bm y}$:
\begin{equation}
dF_{ij}=c\,a_i^{\rm out}(\bm x)a_j^{\rm in}(\bm y)
\,dA_{\bm x}dA_{\bm y},
\label{eq:matching}
\end{equation}
where $c$ is a matching coefficient common to all locations and city pairs. We hold $c$ fixed to isolate how mean intercity flows depend on city populations. Integrating over both cities gives
\begin{equation}
F_{ij}=cQ_i^{\rm out}Q_j^{\rm in}
=GS_i^\alpha S_j^\beta,
\label{eq:target}
\end{equation}
where $G$ collects the coefficients common to all city pairs. The exponent $\alpha$ describes the dependence on the origin city's population, while $\beta$ describes the dependence on the destination city's population.

The difference $\alpha-\beta$ determines the ratio of opposing mean intercity flows:
\begin{equation}
\alpha-\beta=\delta(q_{\rm out}-q_{\rm in}),\qquad
\frac{F_{ij}}{F_{ji}}=
\left(\frac{S_i}{S_j}\right)^{\alpha-\beta}.
\label{eq:symmetry}
\end{equation}
Verbavatz and Barthelemy~\cite{verbavatz2020} find that opposing intercity flows balance on average, while intercity net flows fluctuate over a broad range. We therefore focus on $q_{\rm out}=q_{\rm in}$, so departure and arrival activities have the same density dependence. Equal activity exponents give $\alpha=\beta$ and symmetric mean intercity flows, $F_{ij}=F_{ji}$. Individual realizations of the opposing directional flows can still differ, producing intercity net-flow fluctuations around zero.

For illustration, the cube-root scaling of central density~\cite{lemoy2020,delloye2020} gives $\delta=1/3$ and hence
\begin{equation}
\alpha=\frac{q_{\rm out}+2}{3},\qquad
\beta=\frac{q_{\rm in}+2}{3}.
\end{equation}
Quadratic departure and arrival activities, $q_{\rm out}=q_{\rm in}=2$, then give $\alpha=\beta=4/3$. The value $4/3$ follows from the illustrative choice of activity exponents and is not a measured migration exponent.

\section{Intercity net flows from county-pair correlations}

\subsection{County-pair mean flows and net-flow fluctuations}

Divide city $i$ into $M_i$ counties. For county $a$ occupying area $\Omega_{ia}$, define the population and total departure or arrival activity as
\begin{equation}
S_{ia}=\int_{\Omega_{ia}}n_i\,dA,
\qquad
Q_{ia}^\ell=\int_{\Omega_{ia}}a_i^\ell\,dA,
\label{eq:local_integrals}
\end{equation}
where $\ell\in\{\mathrm{out},\mathrm{in}\}$. Summing over counties gives the city population and total activity:
\begin{equation}
S_i=\sum_a S_{ia},
\qquad
Q_i^\ell=\sum_a Q_{ia}^\ell.
\end{equation}

Let $j_{ia\to jb}$ denote the effective directional flow from county $(i,a)$ to county $(j,b)$. The mean directional flow is
\begin{equation}
\mu_{ia,jb}=\mathbb E[j_{ia\to jb}]
=cQ_{ia}^{\rm out}Q_{jb}^{\rm in}.
\label{eq:local_mean}
\end{equation}
Summing over all county pairs connecting cities $i$ and $j$ gives the mean intercity flow:
\begin{equation}
F_{ij}=\sum_{a,b}\mu_{ia,jb}
=cQ_i^{\rm out}Q_j^{\rm in},
\end{equation}
recovering Eq.~\eqref{eq:target}.

For the remainder of the analysis, we set $q_{\rm out}=q_{\rm in}=q$. Equal departure and arrival activity exponents give symmetric mean flows at both the city and county-pair levels:
\begin{equation}
F_{ij}=F_{ji}=GS_i^hS_j^h,\qquad
\mu_{ia,jb}=\mu_{jb,ia},
\end{equation}
where $h=\alpha=\beta=1+\delta(q-1)$.

Define the county-pair net flow from county $(j,b)$ into county $(i,a)$ as
\begin{equation}
\xi_{ab}^{ij}=j_{jb\to ia}-j_{ia\to jb}.
\end{equation}
Symmetric mean directional flows give $\mathbb E[\xi_{ab}^{ij}]=0$. Verbavatz and Barthelemy~\cite{verbavatz2020} express intercity directional flows as $J_{i\to j}=F_{ij}x_{ij}$, with $\mathbb E[x_{ij}]=1$. The multiplicative representation motivates taking the fluctuation amplitude proportional to the mean directional flow. We assume Gaussian county-pair net flows with
\begin{equation}
\xi_{ab}^{ij}\sim\mathcal N(0,g_{ab}^2),
\qquad
g_{ab}=\lambda\mu_{ia,jb},
\label{eq:local_amplitude}
\end{equation}
where $\lambda$ is the standard deviation of the normalized net flow $\xi_{ab}^{ij}/\mu_{ia,jb}$, common to all county pairs. We treat directional flows as continuous and impose $0<\lambda\le\sqrt{2\pi}$ to allow nonnegative directional flows with the specified means.\footnote{For one county pair, write $\mu=\mu_{ia,jb}$ and $\xi=\xi_{ab}^{ij}$. Define $H=\mu-\mathbb E[|\xi|]/2$ and choose $j_{jb\to ia}=H+\max(\xi,0)$ and $j_{ia\to jb}=H+\max(-\xi,0)$. Both directional flows have mean $\mu$, and the difference is $\xi$. Since $\mathbb E[|\xi|]=\lambda\mu\sqrt{2/\pi}$, the condition $\lambda\le\sqrt{2\pi}$ ensures $H\ge0$ and hence nonnegative directional flows.}

Summing county-pair directional flows gives the intercity directional flow, while summing county-pair net flows gives the intercity net flow:
\begin{equation}
J_{i\to j}=\sum_{a,b}j_{ia\to jb},\qquad
Z_{ij}=J_{j\to i}-J_{i\to j}
=\sum_{a,b}\xi_{ab}^{ij}.
\label{eq:regional_flows}
\end{equation}
The county-pair construction therefore gives $\mathbb E[J_{i\to j}]=F_{ij}$ and $\mathbb E[Z_{ij}]=0$. The identity $Z_{ji}=-Z_{ij}$ ensures that each intercity exchange conserves total population.

\subsection{Correlations between county-pair net flows}

Intercity migration is concentrated in central counties~\cite{reia2022spatial,reia2022modeling}. Fluctuations in central counties may affect several county-pair flows at once. The concentration of migration therefore motivates introducing correlations between county-pair net flows. As a simple model, we assign a common noise to all county pairs connecting a given city pair.

For a fixed city pair, we omit the superscript $ij$ and write
\begin{equation}
\xi_{ab}
=g_{ab}\left[
\sqrt{1-\rho}\,\epsilon_{ab}
+\sqrt{\rho}\,\epsilon_0
\right],
\qquad 0\le\rho\le1.
\label{eq:local_noise}
\end{equation}
The noise variables $\epsilon_{ab}$ and $\epsilon_0$ are mutually independent standard Gaussian variables and are independent of $\rho$. The noise $\epsilon_{ab}$ acts on county pair $(a,b)$ alone, while the noise $\epsilon_0$ acts on all county pairs connecting the two cities.

At a fixed $\rho$, the variance of each county-pair net flow and the covariance between distinct county-pair net flows are
\begin{equation}
\operatorname{Var}(\xi_{ab}\mid\rho)=g_{ab}^2,\qquad
\operatorname{Cov}(\xi_{ab},\xi_{a'b'}\mid\rho)
=\rho g_{ab}g_{a'b'},
\qquad (a,b)\ne(a',b').
\end{equation}
The parameter $\rho$ measures the correlation between distinct county-pair net flows. For every value of $\rho$, each county-pair net flow has the same Gaussian distribution $\mathcal N(0,g_{ab}^2)$. Averaging over different values of $\rho$ therefore leaves each county-pair net-flow distribution unchanged.

The intercity net flow $Z_{ij}=\sum_{a,b}\xi_{ab}$ is Gaussian at a fixed variance $K$:
\begin{equation}
Z_{ij}\mid K\sim\mathcal N(0,K),
\label{eq:conditional_sum}
\end{equation}
where
\begin{equation}
K=(1-\rho)\sum_{a,b}g_{ab}^2
+\rho\left(\sum_{a,b}g_{ab}\right)^2
=A+(B-A)\rho.
\end{equation}
Using $g_{ab}=\lambda\mu_{ia,jb}$ and $\sum_{a,b}\mu_{ia,jb}=F_{ij}$ gives
\begin{equation}
A=\lambda^2\sum_{a,b}\mu_{ia,jb}^2,\qquad
B=\lambda^2F_{ij}^2.
\label{eq:variance_bounds}
\end{equation}
Independent county-pair net flows ($\rho=0$) give the minimum variance $A$, while fully correlated county-pair net flows ($\rho=1$) give the maximum variance $B$.

The variance bounds depend on how activity is distributed across counties. Define county $a$'s share of the total activity in city $i$ as
\begin{equation}
w_{ia}
=\frac{Q_{ia}^{\rm out}}{Q_i^{\rm out}}
=\frac{Q_{ia}^{\rm in}}{Q_i^{\rm in}},
\qquad
\sum_a w_{ia}=1.
\end{equation}
The departure and arrival activity shares are equal because departure and arrival activities have the same density dependence. Equation~\eqref{eq:local_mean} gives
\begin{equation}
\mu_{ia,jb}=F_{ij}w_{ia}w_{jb}.
\end{equation}
Substituting the county-pair mean flows into Eq.~\eqref{eq:variance_bounds} yields
\begin{equation}
\frac{B}{A}
=\frac{1}{\left(\sum_a w_{ia}^2\right)
\left(\sum_b w_{jb}^2\right)}
=M_{{\rm eff},i}M_{{\rm eff},j},
\label{eq:effective_count}
\end{equation}
where
\begin{equation}
M_{{\rm eff},i}=\frac{1}{\sum_a w_{ia}^2},
\qquad
1\le M_{{\rm eff},i}\le M_i.
\end{equation}
The quantity $M_{{\rm eff},i}$ measures the effective number of active counties in city $i$. Activity spread across many counties with comparable shares gives a larger effective number, while activity concentrated in a few counties gives a smaller effective number. The product of the effective numbers in the two cities determines the variance ratio $B/A$.

\subsection{A statistical construction of the variance distribution}

At a fixed correlation $\rho$ between county-pair net flows, the intercity net flow is Gaussian with variance $K(\rho)=A+(B-A)\rho$. Averaging the conditional Gaussian density over the correlation density $p_\rho(\rho)$ gives
\begin{equation}
p_{ij}(z)
=\int_0^1
\frac{\exp[-z^2/(2K(\rho))]}{\sqrt{2\pi K(\rho)}}
p_\rho(\rho)\,d\rho.
\end{equation}
For $A=B$, the variance is fixed and the intercity net-flow distribution remains Gaussian. For $A<B$, the intercity net-flow distribution depends on the correlation density $p_\rho$, or equivalently on the variance density $p_K$. We construct a variance density that produces an intermediate power law when the variance range is sufficiently broad.

Following the squared-Gaussian construction in superstatistics~\cite{beck2001}, we define
\begin{equation}
\tau^{(0)}=d\sum_{\ell=1}^{m}Y_\ell^2,
\qquad
Y_\ell\sim\mathcal N(0,1),
\quad d>0,
\label{eq:squared_gaussians}
\end{equation}
where $m$ is a positive integer. The Gaussian variables $Y_\ell$ are mutually independent and independent of the noise variables in Eq.~\eqref{eq:local_noise}. The scaled sum of squares $\tau^{(0)}$ follows the gamma density
\begin{equation}
p_\tau^{(0)}(\tau)=\frac{b^\nu}{\Gamma(\nu)}
\tau^{\nu-1}e^{-b\tau},
\qquad
\tau>0,\quad
\nu=\frac m2,\quad b=\frac1{2d}.
\label{eq:gamma_precision}
\end{equation}

To keep the variance within $[A,B]$, we restrict the gamma density to $[1/B,1/A]$ and normalize the restricted density. We denote the resulting variable by $\tau$ and define the variance as $K=1/\tau$. Changing variables from $\tau$ to $K$, with $|d\tau/dK|=K^{-2}$, gives
\begin{equation}
p_K(K)=CK^{-\nu-1}e^{-b/K},
\qquad A\le K\le B,\qquad
C^{-1}=\int_A^B K^{-\nu-1}e^{-b/K}\,dK.
\label{eq:variance_density}
\end{equation}
The variance therefore follows an inverse-gamma density restricted to $[A,B]$, with normalization constant $C$.

The relation $\rho=(K-A)/(B-A)$ then gives the correlation density:
\begin{equation}
p_\rho(\rho)=(B-A)p_K\!\left(A+(B-A)\rho\right),
\qquad 0\le\rho\le1.
\end{equation}

\subsection{An intermediate power law and a Gaussian cutoff}

Substituting the variance density in Eq.~\eqref{eq:variance_density} into the Gaussian mixture gives
\begin{equation}
p_{ij}(z)=\frac{C}{\sqrt{2\pi}}
\int_A^B K^{-\nu-3/2}e^{-(b+z^2/2)/K}\,dK.
\label{eq:mixture}
\end{equation}
To identify the power-law range, define $v=b+z^2/2$ and change variables to $u=v/K$:
\begin{equation}
p_{ij}(z)=\frac{C}{\sqrt{2\pi}}v^{-\nu-1/2}
\int_{v/B}^{v/A}u^{\nu-1/2}e^{-u}\,du.
\label{eq:transformed_mixture}
\end{equation}
For fixed $\nu$ and $A\ll v\ll B$, the lower integration limit is close to zero and the upper integration limit is large. Extending the integration limits to zero and infinity gives
\begin{equation}
\int_{v/B}^{v/A}u^{\nu-1/2}e^{-u}\,du
\simeq \int_0^\infty u^{\nu-1/2}e^{-u}\,du
=\Gamma(\nu+1/2).
\end{equation}
The intercity net-flow density therefore satisfies
\begin{equation}
p_{ij}(z)\simeq
\frac{C\Gamma(\nu+1/2)}{\sqrt{2\pi}}
(b+z^2/2)^{-\nu-1/2}.
\end{equation}

Choose $d$ large enough that $b=1/(2d)\ll A$. Over the range
\begin{equation}
A\ll z^2/2\ll B,
\end{equation}
the approximation $b+z^2/2\simeq z^2/2$ gives
\begin{equation}
\boxed{
p_{ij}(z)\propto |z|^{-\gamma},
\qquad \gamma=2\nu+1.
}
\label{eq:tail}
\end{equation}
The squared-Gaussian construction gives $\nu=m/2$ and hence $\gamma=m+1$. A broad power-law range requires a broad variance range, $B\gg A$.

The finite maximum variance $B$ produces a Gaussian cutoff. As $|z|\to\infty$ with $A$, $B$, $b$, and $\nu$ fixed,
\begin{equation}
p_{ij}(z)\propto |z|^{-2}e^{-z^2/(2B)}.
\end{equation}
The intercity net-flow distribution therefore follows an intermediate power law with a Gaussian cutoff. The dependence of the tail on the variance distribution is consistent with general analyses of superstatistics~\cite{touchette2005}.

\subsection{A dynamical construction of the variance distribution}

A drift toward a reference value and noise proportional to the evolving variable can generate an inverse-gamma distribution, as in mean-field exchange models~\cite{bouchaud2000}. To construct the variance density in Eq.~\eqref{eq:variance_density}, we let the variance $K$ evolve according to the It\^o equation
\begin{equation}
dK=\kappa(K_0-K)\,dt+\sigma K\,dW,
\label{eq:variance_SDE}
\end{equation}
where $\kappa,\sigma>0$, $A<K_0<B$, and $W$ is a standard Brownian motion. The drift drives $K$ toward the reference variance $K_0$, while the noise amplitude increases in proportion to $K$. Reflecting boundaries at $A$ and $B$ keep the variance within $[A,B]$ and impose zero probability current at both boundaries.

The variance density $p_K(K,t)$ obeys the Fokker--Planck equation
\begin{equation}
\partial_t p_K
=-\partial_K\!\left[\kappa(K_0-K)p_K\right]
+\frac{\sigma^2}{2}\partial_K^2(K^2p_K).
\end{equation}
Writing the Fokker--Planck equation as $\partial_t p_K=-\partial_K J_K$ gives the probability current
\begin{equation}
J_K(K,t)
=\kappa(K_0-K)p_K
-\frac{\sigma^2}{2}\partial_K(K^2p_K).
\end{equation}
At stationarity, $\partial_K J_K=0$, so the probability current is constant. The reflecting boundary conditions $J_K(A)=J_K(B)=0$ therefore give zero probability current throughout $[A,B]$:
\begin{equation}
\kappa(K_0-K)p_K
-\frac{\sigma^2}{2}\partial_K(K^2p_K)=0.
\label{eq:stationary_current}
\end{equation}
Solving Eq.~\eqref{eq:stationary_current} gives the stationary variance density
\begin{equation}
p_K(K)\propto
K^{-2-2\kappa/\sigma^2}
\exp\!\left(-\frac{2\kappa K_0}{\sigma^2K}\right),
\qquad A\le K\le B.
\end{equation}
The stationary density is the restricted inverse-gamma density in Eq.~\eqref{eq:variance_density}, with
\begin{equation}
\nu=1+\frac{2\kappa}{\sigma^2},
\qquad
b=\frac{2\kappa K_0}{\sigma^2}.
\end{equation}
The dynamical construction therefore requires $\nu>1$ and $b/(\nu-1)=K_0\in(A,B)$. The reflecting boundaries can shift the stationary mean away from the reference variance $K_0$.

For $b=2\kappa K_0/\sigma^2\ll A$ and $B\gg A$, averaging the conditional Gaussian density over the stationary variance density gives
\begin{equation}
p_{ij}(z)\propto |z|^{-\gamma},
\qquad
\gamma=3+\frac{4\kappa}{\sigma^2},
\end{equation}
over the range $A\ll z^2/2\ll B$, as follows from Eq.~\eqref{eq:tail}. The finite maximum variance $B$ produces a Gaussian cutoff beyond the power-law range.

As $\sigma\to0$ with all other parameters fixed, the stationary variance distribution concentrates at $K_0$. The intercity net-flow distribution then approaches a Gaussian with variance $K_0$.

\section{Discussion}

We construct mean intercity flows and intercity net-flow distributions from the same county-pair exchanges. Density-dependent departure and arrival activities determine how mean intercity flows scale with the populations of the origin and destination cities. We assume Gaussian county-pair net flows with standard deviations proportional to the corresponding mean directional flows. Changing the correlations between county-pair net flows preserves each county-pair Gaussian distribution but changes the variance of the intercity net flow.

The mean intercity flow and the fluctuation amplitude $\lambda$ set the maximum variance $B$, while county activity shares determine the variance ratio $B/A=M_{{\rm eff},i}M_{{\rm eff},j}$. Activity spread across many counties with comparable shares allows a broad variance range. Independent county-pair fluctuations partly cancel, while positively correlated county-pair fluctuations tend to reinforce one another. A separate assumption specifies how frequently different variances occur within $[A,B]$.

We present two constructions of a restricted inverse-gamma variance distribution. The statistical construction generates the inverse variance from a sum of squared Gaussian variables. The Gaussian variables $Y_\ell$ may represent independent socioeconomic factors with different effects on different county pairs. Stronger fluctuations in the socioeconomic factors may make county-pair net flows move less closely together, reducing correlations and intercity net-flow variance.

The dynamical construction generates the variance through a drift toward a reference variance and noise proportional to the variance. The drift may represent a return toward a typical level of migration variability. The proportional noise may represent greater sensitivity to socioeconomic shocks when migration variability is high. Shocks of the same magnitude then produce larger changes in variance at higher variance levels. The socioeconomic interpretations of both constructions require empirical evaluation.

For $b\ll A$ and $B\gg A$, averaging the conditional Gaussian density over the variance distribution produces an intermediate power law with exponent $\gamma=2\nu+1$ and a Gaussian cutoff. Heavy-tailed intercity net-flow distributions can therefore arise from varying correlations even when every county-pair net-flow distribution remains Gaussian.

The model suggests three directions for further study. First, network structure affects collective behavior in spin models~\cite{dorogovtsev2008critical,jang2015ashkin,kim2025ashkin,kim2026heterogeneous} and transport in random walks~\cite{hwang2012first,hwang2013origin,masuda2017random,kim2025spatial}. Introducing a migration network in place of connections between all county pairs would allow us to examine how hubs and heterogeneous connections affect correlations between county-pair net flows and intercity net-flow distributions.

Second, inertia and time delays affect dynamics in coupled oscillators~\cite{yeung1999delay,kim2025cluster,kim2026paths}, while memory affects transport in networks~\cite{rosvall2014memory}. Introducing an explicit persistence time for correlations between county-pair net flows would allow us to examine how intercity net-flow distributions depend on the observation period relative to the persistence time.

Third, positive feedback can amplify local disturbances in threshold models~\cite{watts2002cascades} and models of cascading failures~\cite{motter2002cascade,jhun2023prediction,park2025optimal}. Introducing feedback between population redistribution and migration flows would allow us to examine how migration changes county population densities, how population densities change departure and arrival activities, and how departure and arrival activities change subsequent flows. The feedback model could identify conditions under which migration amplifies net-flow fluctuations and produces large population changes across cities.

\bibliographystyle{apsrev4-2}
\bibliography{references}
\end{document}